\documentclass[fdp,a4paper,fleqn%
]{w-art}
\usepackage{times,cite,w-thm}
\theoremstyle{plain}

\theoremstyle{definition}

\usepackage[]{graphicx}
\usepackage{comment}
\begin{document}
\DOIsuffix{theDOIsuffix}
\Volume{55}
\Month{01}
\Year{2007}
\pagespan{1}{}
\Receiveddate{XXXX}
\Reviseddate{XXXX}
\Accepteddate{XXXX}
\Dateposted{XXXX}



\title[Superfluorescence from Quantum Wells]{Generation of Superfluorescent Bursts from a Fully Tunable Semiconductor Magneto-plasma}


\author[Noe]{G. Timothy Noe II\inst{1}}
\address[\inst{1}]{Departments of Electrical and Computer Engineering and Physics and Astronomy, Rice University, Houston, Texas 77005, USA}

\author[Kim]{Ji-Hee Kim\inst{1}}

\author[Lee]{Jinho Lee\inst{2}}
\address[\inst{2}]{Department of Physics, University of Florida, Gainesville, Florida 32611, USA}

\author[Jho]{Young-Dahl Jho\inst{3}}
\address[\inst{3}]{Department of Information and Communications, Gwangju Institute of Science and Technology, Gwangju 500-712, Republic of Korea}

\author[Wang]{Yongrui Wang\inst{4}}
\address[\inst{4}]{Department of Physics and Astronomy, Texas A\&M University, College Station, Texas 77843, USA}

\author[W\'{o}jcik]{Aleksander K. W\'{o}jcik\inst{4}}

\author[McGill]{Stephen A. McGill\inst{5}}
\address[\inst{5}]{National High Magnetic Field Laboratory, Florida State University, Tallahassee, Florida 32310, USA}

\author[Reitze]{David H. Reitze\inst{2}}

\author[Belyanin]{Alexey A. Belyanin\inst{4}}

\author[Kono]{Junichiro Kono\inst{1,}%
  \footnote{Corresponding author\quad E-mail:~\textsf{kono@rice.edu},
            Phone: +00\,1\,713\,348\,2209,
            Fax: +00\,1\,713\,348\,5686}}

\begin{abstract}
  Quantum particles sometimes cooperate to develop a macroscopically ordered state with extraordinary properties.  Superconductivity and Bose-Einstein condensation are examples of such cooperative phenomena where macroscopic order appears {\em spontaneously}.  Here, we demonstrate that such an ordered state can also be obtained in an optically excited semiconductor quantum well in a high magnetic field.  When we create a dense electron-hole plasma with an intense laser pulse, after a certain delay, an ultrashort burst of coherent radiation emerges.  We interpret this striking phenomenon as a manifestation of superfluorescence (SF), in which a macroscopic polarization spontaneously builds up from an {\em initially incoherent ensemble of excited quantum oscillators} and then decays abruptly {\em producing giant pulses of coherent radiation}.  SF has been observed in atomic gases, but the present work represents the first observation of SF in a solid-state setting.  While there is an analogy between the recombination of electron-hole pairs and radiative transitions in atoms, there is no {\it a priori} reason for SF in semiconductors to be similar to atomic SF.  This is a {\em complex many-body system with a variety of ultrafast interactions}, where the decoherence rates are at least 1,000 times faster than the radiative decay rate, an unusual situation totally unexplored in previous atomic SF studies.  We show, nonetheless, that collective many-body coupling via a common radiation field does develop under certain conditions and leads to SF bursts.  The solid-state realization of SF resulted in an unprecedented degree of controllability in the generation of SF, opening up opportunities for both fundamental many-body studies and device applications.  We demonstrate that the intensity and delay time of SF bursts are fully tunable through an external magnetic field, temperature, and pump laser power.
\end{abstract}
\maketitle                   





\section{Introduction}

Currently, considerable resurgent interest exists in the concept of superradiance (SR), i.e., accelerated relaxation of excited dipoles due to cooperative spontaneous emission, first proposed by Dicke in 1954~\cite{Dicke54PR}.  Recent authors have discussed SR in diverse contexts, including cavity quantum electrodynamics~\cite{ScullySvidzinsky09Science}, quantum phase transitions~\cite{BaumannetAl10Nature}, and plasmonics~\cite{SonnefraudetAl10ACSNano,Martin-CanoetAl10NL}.  At the heart of these various experiments lies the coherent coupling of constituent particles to each other via their radiation field that cooperatively governs the dynamics of the whole system.  In the most exciting form of SR, called superfluorescence (SF)~\cite{BonifacioLuiato75PRA,BKK92,BelyaninetAl92,BelyaninetAl97QSO,BelyaninetAl98QSO}, macroscopic coherence spontaneously builds up out of an initially incoherent ensemble of excited dipoles and then decays abruptly.

Here, we review a series of studies that we have recently performed on an ultradense electron-hole plasma in a semiconductor in a magnetic field~\cite{JhoetAl06PRL,JhoetAl10PRB}, culminating in the direct, time-domain observation of SF bursts~\cite{NoeetAl12NP}.  While optically excited semiconductors have attracted continuing interest for many years through endless discoveries associated with high-density electron-hole pairs, excitons, and polaritons~\cite{ChemlaShah01Nature,ButovetAl02Nature,ButovetAl02Nature2,KasprzaketAl06Nature,LietAl06PRL,BalilietAl07Science,TurnerNelson10Nature,DengetAl10RMP}, this is the first direct observation of SF in a semiconductor and offers numerous future opportunities for further studies and device applications.  We observed intense, delayed pulses, or bursts, of coherent radiation from highly photo-excited semiconductor quantum wells (QWs) via time-resolved photoluminescence (TRPL), while, at the same time, we observed a sudden decrease in population from total inversion to zero via ultrafast pump-probe spectroscopy.  The SF bursts were delayed in time by tens of picoseconds with respect to the femtosecond excitation pulse, and they appeared only when the excitation powers and magnetic fields were above some critical values and the temperature was low enough.  We performed theoretical simulations based on the relaxation and recombination dynamics of ultrahigh-density electron-hole pairs in a quantizing magnetic field, which successfully captured the salient features of the experimental observations.

\section{Experimental Methods}

We performed time-integrated photoluminescence (TIPL) spectroscopy~\cite{JhoetAl06PRL,JhoetAl10PRB}, time-resolved pump-probe spectroscopy, and TRPL spectroscopy~\cite{NoeetAl12NP} measurements on an In$_{0.2}$Ga$_{0.8}$As multiple QW sample as a function of magnetic field, temperature, and under a variety of excitation conditions.  The QW sample consisted of fifteen 8-nm In$_{0.2}$Ga$_{0.8}$As wells separated by 15-nm GaAs barriers~\cite{JhoetAl05PRB}.  The GaAs substrate was transparent to light with photon energies corresponding to the lowest energy states of the QWs, allowing for transmission measurements without etching away the substrate.  Furthermore, the strain present in the sample lifted the degeneracy between the lowest heavy-hole subband ($H_1$) and lowest light-hole subband ($L_1$), resulting in little energy level mixing behavior between the $E_1 H_1$ and $E_{1}L_{1}$ interband transitions of the QWs~\cite{JhoetAl05PRB}.  All measurements were performed in the Fast Optics facility at the National High Magnetic Field Laboratory (NHMFL) in Tallahassee, Florida.

\begin{figure}
\begin{center}
\includegraphics[scale = 0.62]{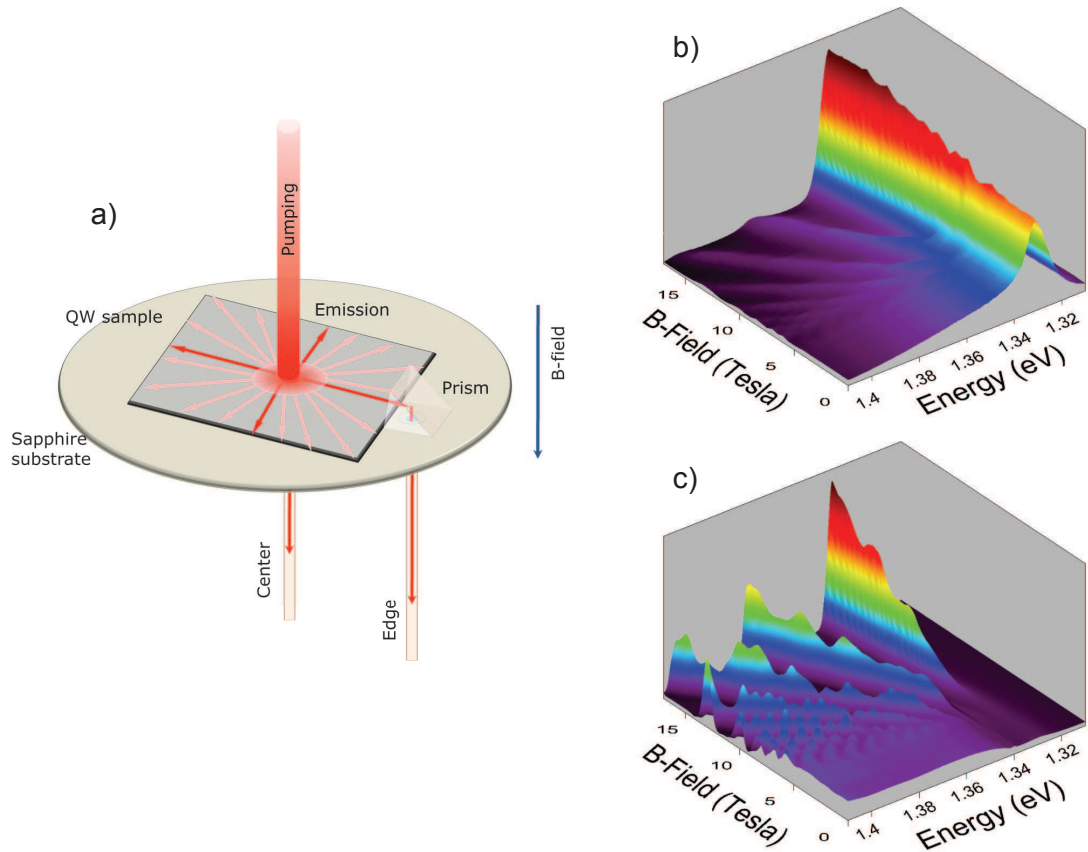}
\caption{\textbf{a)} Optical setup. Adapted from \cite{NoeetAl12NP}. \textbf{b)} Magnetic field dependence for the time-integrated PL measured from the center collection fiber directly behind the excitation spot.  \textbf{c)} Magnetic field dependence for the time-integrated PL measured from the edge collection fiber after total internal reflection with a micro-prism.}
\label{setup}
\end{center}
\end{figure}

\medskip

{\it Time-Integrated Magneto-photoluminescence Spectroscopy:} TIPL measurements were done using either a 31~Tesla dc resistive magnet or a 17.5~Tesla superconducting magnet with direct access to excitation by using a 1~kHz Ti:Sapphire-based regenerative amplifier laser system to deliver up to $\mu$J pulses normally incident to the QWs in the Faraday geometry, where the magnetic field and optical excitation beam are parallel.  Multimode optical fibers were used to collect emission either from the opposite side of the sample from excitation or on the cleaved edge after total internal reflection from a micro-prism located at the edge to redirect the in-plane emission.  See Fig.~\ref{setup}(a).  The spectra of the light collected in these manners was measured with a CCD spectrometer.  Spectra were taken by averaging over many pulses or by doing single shot measurements with the use of a Pockels cell external to the laser.

\medskip

{\it Pump-Probe Spectroscopy:} Ultrafast pump-probe spectroscopy measurements in a transmission geometry were made by using traditional delay stage techniques, where the amplified Ti:Sapphire laser beam pumped the sample and an optical parametric amplifier's (OPA) output [tuned to the photon energy corresponding to specific Landau level (LL) transitions] probed the population dynamics of electron-hole pairs in the LLs.  The pump beam was modulated using an optical chopper and the probe light, after collection with the center collection fiber and spectrally filtered to a $\sim$2~nm bandwidth centered at the peak of the in-plane emission, was measured with a silicon photodiode, and the resulting electronic signal was sampled with a lock-in amplifier using the modulation frequency of the chopper as the reference.  Data was taken for the lowest three LL transitions as a function of magnetic field and temperature.

\medskip

{\it Time-Resolved Photoluminescence Spectroscopy:} TRPL measurements were done by using a 2~ps resolution streak camera with a grating monochromator before the entrance slit to spectrally resolve the PL.  The amplified Ti:Sapphire laser beam was, again, used to pump the sample for high excitation measurements.  The PL was collected with graded-index fibers for both center and edge collection to reduce the dispersion as the PL travels from inside the magnet to the entrance of the spectrometer.  The OPA output (tuned to various wavelengths and measured with the streak camera immediately proceeding the PL measurement) was used as a calibration tool to correct the timing of the PL based on the fact that different wavelengths have different travel times through the fiber and in the monochromator.  By measuring the OPA under the same circumstances as the PL, our temporal resolution of the system was determined to be $\sim$20~ps.

\section{Experimental Results}


The time-integrated measurements show very different emission properties depending on whether the PL is collected from the center or from the edge, as shown in Figs.~\ref{setup}{\bf b)} and \ref{setup}{\bf c)}, respectively.  With increasing magnetic field, both sets of data show emission from multiple LL transitions arising from the $E_{1}H_{1}$ transition.  In the center collection data, by far the most dominant feature is the PL from the lowest energy transition in the well, the (00) LL transition, which increases gradually with increasing magnetic field.  Emission from other LLs can be seen, but the emission is rather weak and arises as small contributions on the high energy tail of the (00) LL PL.  In the edge collection data, on the other hand, there is a dramatic increase in PL with increasing magnetic field.  At the lower fields, emission from the (00) LL is evident as a small contribution on the side of higher energy emission.  With increasing magnetic field, the (00) LL PL increases dramatically, and the emission from other LLs also becomes very bright.  In contrast to the center collected PL, at high magnetic fields, the PL collected from the edge is much brighter, the peaks are much sharper, and they are completely spectrally separated.  Also, a magnetic field dependent intensity modulation can be seen for the first few LLs.

\begin{figure}
\begin{center}
\includegraphics[scale=1.15]{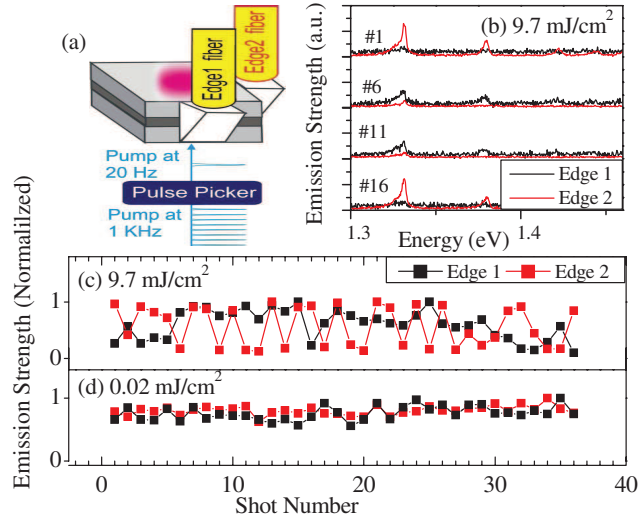}
\caption{{\bf a)}~Optical setup for shot-to-shot time-integrated PL measurements.  {\bf b)}~Representative single-shot spectra from two edges.  Normalized emission strength for the (00) LL at {\bf c)}~high fluence, showing anti-correlation behavior between the two edges, and {\bf d)}~low fluence, showing simultaneous detection consistent with isotropic emission. Figure from \cite{JhoetAl06PRL}.}
\label{YD}
\end{center}
\end{figure}

In order to investigate the directionality of the emission, single-shot measurements were done using two edge collection fibers as illustrated in Fig.~\ref{YD}.  Time-integrated spectra for both emission directions were taken simultaneously on a shot-to-shot basis.  The normalized emission strength for the (00) LL versus shot number is shown under two conditions: 9.7~mJ/cm$^2$ with a 0.5~mm spot size and 0.02~mJ/cm$^2$ with a 3~mm spot size.  The results show a strong anti-correlation between the emission strengths in the two directions under the higher pump fluence and a correlation under the lower pump fluence.  These results indicate directional emission in the case of high pump fluence but with the directionality randomly changing from shot to shot.

\begin{figure}
\includegraphics[scale=0.9]{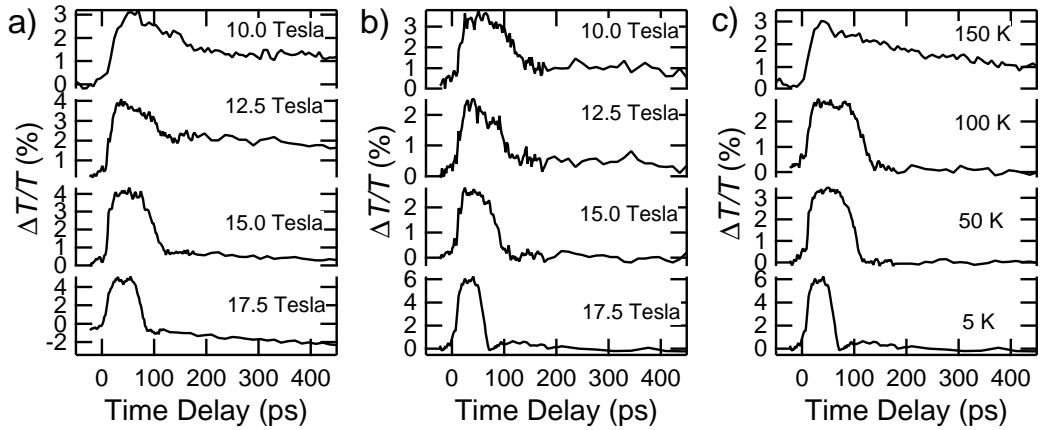}
\caption{Magnetic field dependent time-resolved pump-probe differential transmission data for probing the {\bf a)}~(11) and {\bf b)}~(22) LL at 5~K.  {\bf c)}~Temperature dependent pump-probe differential transmission data for the (22)~LL at 17.5~Tesla.  Adapted from \cite{NoeetAl12NP}.}
\label{PP}
\end{figure}

In order to measure the population dynamics of electrons and holes in the LLs, pump-probe differential transmission measurements were done as a function of magnetic field at 5~K and as a function of temperature at 17.5~T, as shown in Fig.~\ref{PP}.  With increasing magnetic field, the differential transmission data shows a transition from an exponential-like decay at lower magnetic fields to a sudden drop to zero at the highest magnetic field for the (11)~LL, as shown in Fig.~\ref{PP}{\bf a)}.  The data for the (22)~LL, shown in Fig.~\ref{PP}{\bf b)}, also shows a sudden drop to zero, but the abrupt decrease happens at earlier time delays.  Also, the emergence of an accelerated decrease happens at lower magnetic fields for the (22)~LL than for the (11)~LL transition.  Finally, as shown in Fig.~\ref{PP}{\bf c)}, decreasing the temperature has a similar effect as increasing the magnetic field in that the sudden decrease emerges at lower temperatures and happens at earlier time delays for the lowest temperature.

\begin{figure}[b]
\begin{center}
\includegraphics[scale=0.8]{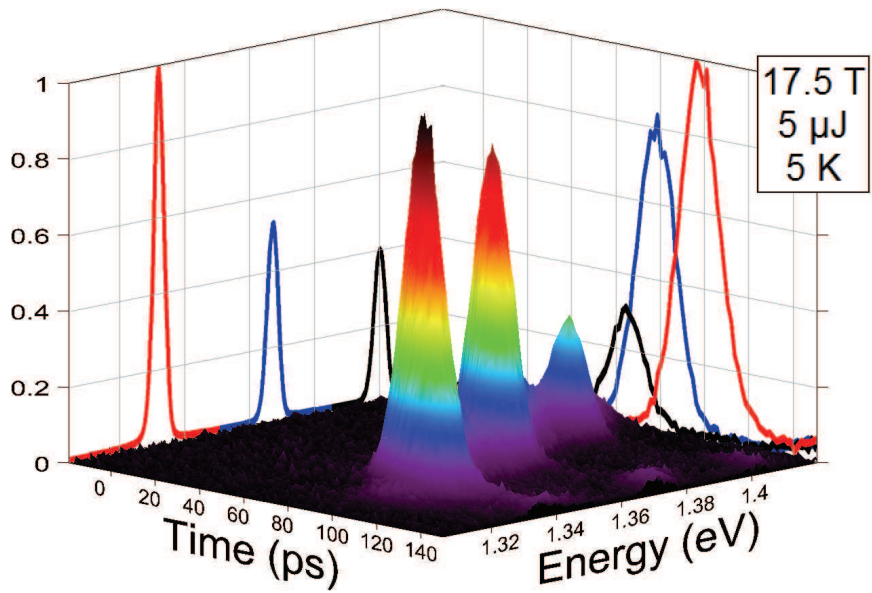}
\caption{Spectrally and temporally resolved PL at 17.5~T, 5~K, and with 5~$\mu$J pump pulse energy.  The left hand panel shows the time-integrated spectra and the right hand panel shows spectrally selected time-resolved slices for the (00), (11), and (22)~LLs with color codes of red, blue, and black respectively.  Adapted from \cite{NoeetAl12NP}.}
\label{TRPL}
\end{center}
\end{figure}

\begin{figure}
\begin{center}
\includegraphics[scale=0.7]{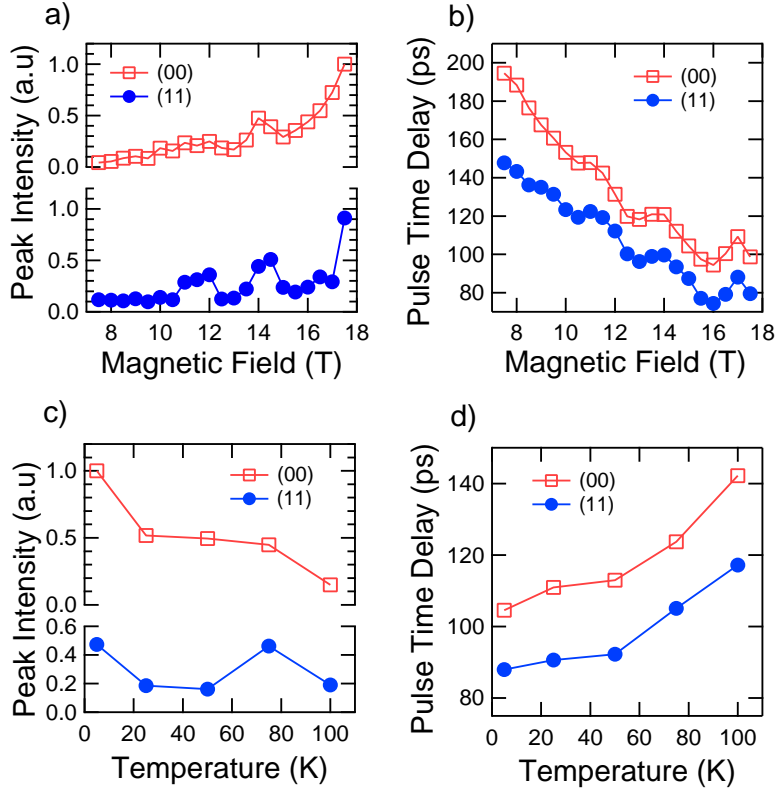}
\caption{{\bf a)} Peak intensity and {\bf b)} pulse time delay as a function of magnetic field for the (00) and (11) LL SF bursts.  {\bf c)}~Peak intensity and {\bf d)}~pulse time delay as a function of temperature for the (00) and (11) LL SF bursts.}
\label{fits}
\end{center}
\end{figure}

The TRPL was recorded for various magnetic fields at the lowest temperature, various temperatures at the highest magnetic field, and various excitation pulse energies at the highest magnetic field and lowest temperature.  Figure \ref{TRPL} shows an example of the TRPL at 17.5~T, 5~K, and with 5~$\mu$J pump pulse energy, in which large bursts of radiation after some time delay are observed, discrete both in energy and time.  The large bursts occur in order of decreasing photon energy where the (22)~LL emission occurs first and then the (11)~LL emission followed by the (00)~LL emission.  Figure \ref{fits} shows the effect of increasing the magnetic field or decreasing the temperature on the intensity and time delay for the SF bursts at the photon energies corresponding to the (00) and (11)~LL transitions.  The magnetic field dependent data was taken at the lowest temperature, 5~K, and the temperature dependent data was taken at the highest magnetic field, 17.5~T.  Both increasing the magnetic field or decreasing the temperature show a similar trend in that the peak intensity increases and the pulse time happens at an earlier time delay.  On top of this trend, there is a magnetic field dependent oscillation in these two quantities.

\section{Discussion: Theory of SF from a Semiconductor Magneto-plasma}

The salient features of SF are similar for any ensemble of dipole oscillators: atoms, excitons, impurity ions, or electron-hole pairs.  An incoherently prepared ensemble of excited oscillators interacts via the exchange of spontaneously emitted photons.  At sufficiently high density $N$ of ``atoms'' and sufficiently low dephasing rate, this exchange leads to the effective mutual phasing of the optical oscillations.  As a result, a macroscopic optical polarization with an amplitude $\propto N$ spontaneously forms, arising from quantum fluctuations, with a characteristic formation time, or pulse delay time, $t_d$.  The resulting macroscopic polarization decays by emitting a coherent radiation pulse of duration, $t_p$, shorter than the times of incoherent spontaneous emission and phase relaxation in the medium: $t_p, t_d < T_1, T_2$.  This is a distinctive feature of the phenomenon.  In addition, SF is fundamentally a stochastic process: the optical polarization and the electromagnetic field grow from initially incoherent quantum noise to a macroscopic level.  Thus, SF is intrinsically random: for identical preparation conditions, initial microscopic fluctuations get exponentially amplified and result in the macroscopic pulse-to-pulse fluctuations in the delay time $t_d$, electric field polarization, and the direction of the emitted pulse.

The key to achieving SF is to provide a high enough spatial and spectral density of inverted dipole oscillators such that the growth rate of the polarization exceeds the dephasing rate ($1/T_2$).  In semiconductors, the rate of the optical polarization decay is typically 1\,ps$^{-1}$ or greater, whereas the maximum density of electron-hole pairs within the bandwidth $\Delta\omega$ of the SF pulse is limited by the density of states $\rho(E)$: $N_{\rm max} \sim \rho(E) \hbar \Delta \omega$.  It is this combination of a limited gain and fast decoherence that makes it difficult to achieve the threshold for SF~\cite{BKK92,BelyaninetAl97QSO}.  One possible way to overcome these limitations is to use QWs placed in strong magnetic fields~\cite{BKK92}. The combination of reduced dimensionality and the magnetic field fully quantizes the semiconductor into an atomic-like system with a series of Landau levels (LLs), thus strongly increasing the density of states (DOS) and enhancing the oscillator strength. In addition, a complete quantum confinement and filling of all available electron states at the lowest LLs are expected to suppress the dephasing rate of the optical polarization.  In our experiments, the initial pumping conditions are chosen so as to ensure the growth of coherence from an incoherent state by exciting the initial electron-hole plasma high in the bands with an excess energy of 150\,meV above the GaAs barrier bandgap.  The energy difference between the excitation energy and the (00) LL in our QWs is greater than 270\,meV, requiring many scattering events to relax into the QW LLs.  The resulting degenerate electron-hole system residing in the lowest QW subband is thus initially completely incoherent.

The discrimination between SF and other recombination emission regimes in semiconductors is convenient to formulate in terms of a linear initial-value problem, assuming constant initial degenerate populations of carriers created by pumping and subsequent relaxation~\cite{BelyaninetAl97QSO,JhoetAl10PRB}. After finding the linear susceptibility, one can solve the dispersion and find the complex frequency $\omega(k)$ as a function of a real wavenumber $k$.  The instability, or exponential growth of the field and polarization with time, develops within the spectral interval of wavenumbers where Im$[\omega(k)] > 0$. Obviously, for instability of any kind, one needs positive population inversion  with respect to stimulated recombination vs.~absorption processes at the photon frequency.

Let us consider an idealized situation when all scattering rates and transition frequencies are the same for all electron-hole pairs or excitons on a given LL, similar to a homogeneously broadened active medium.  Note that each spin-split LL has a degeneracy of $N_{\rm 2D} = eB/(2\pi\hbar c) \simeq 2.4\times 10^{10} B(T)$~cm$^{-2}$, and population inversion $\Delta N > 0$ can be defined as the difference between occupied and unoccupied states per unit area. It is related to the density of excitons $N_{\rm exc}$ as $\Delta N = 2 N_{\rm exc} - 1$.   Now, the key parameter governing  different regimes of recombination is the cooperative frequency, $\Omega_c$, which determines the coupling strength between the field and the optical polarization as introduced in previous studies of SF in semiconductors~\cite{BelyaninetAl92,BelyaninetAl98QSO}:
\begin{equation} 
\label{coop}
\Omega_c = \sqrt{\frac{16 \pi^2 d^2 \Delta N \Gamma c}{\hbar \mu^2
\lambda L_{\rm QW}}}.
\end{equation}
Here, $L_{\rm QW}$ is the total width of the QWs (i.e., the sum of widths of all wells in the sample), $\mu$ the refractive index, $d$ the dipole matrix element of the interband optical transition, and $\lambda$ is the wavelength of the radiated field.  When the inversion $\Delta N$ is small or negative, field oscillations decay with time, and one can have only spontaneous recombination emission with power $\propto N_{\rm exc}/t_{\rm sp}$, and the characteristic timescale $t_{\rm sp} \sim 1$\,ns.  With increasing inversion, the modal gain may become higher than losses, and oscillations of the field and the polarization will grow with time.

There are two basic regimes of instability, depending on the ratio between the values of $\Omega_c$ and incoherent relaxation times.  For low inversion density and fast decoherence, when $\Omega_c \ll \gamma = 1/T_2$, one can have amplified spontaneous emission (ASE), provided that the electromagnetic field decay rate $\alpha $ is low enough.  This is a regime of a one-pass amplifier.  Its growth rate
\begin{equation} 
\label{lase}
{\rm Im}(\omega) = g_{\rm ASE} \approx \frac{\omega_c^2 \gamma}{
4\{\gamma^2 + (c/\mu)^2 (k-k_0)^2 \}} -c \alpha/\mu \ll \gamma
\end{equation}
is much slower than the dephasing rate $\gamma$. Equation (\ref{lase}) is obtained via first-order perturbation theory by substituting $\omega \approx ck/\mu$ in the expression for the linear susceptibility  $\chi(\omega)$. This approximation is equivalent to calculating the net rate of stimulated transitions using Fermi's golden rule.  When the gain increases to the value such that $g_{\rm ASE}L \mu/c \gg 1$, amplification proceeds in the saturated regime, which is sometimes called superluminescence (SL).  Here, $L$ is the length of an active medium in the propagation direction.  The duration of the SL pulse in a saturated amplifier is $t_{\rm SL} \sim L
\mu/c$~\cite{Benedict96}.  We emphasize that all timescales for the above processes of ASE or SL are longer than the dephasing time $T_2$.  Under such conditions, the optical polarization has a small amplitude and adiabatically follows the time dependence of the electromagnetic field. This implies low mutual coherence of dipole oscillations of individual electron-hole pairs.

Cooperative recombination, or SF, develops when the growth rate of the polarization is larger than the decay rate $\gamma$.  The instability results in the spontaneous formation of a large-amplitude coherent macroscopic polarization from initially incoherent oscillations of individual electron-hole dipole moments.  Here, {\em spontaneous} means that the polarization is not created by some external coherent laser field. The necessary condition for the instability is $\Omega_c > 2\gamma$.  When the electromagnetic field decay rate $\alpha$ is very high such that $\gamma < \Omega_c /2 < \alpha$, cooperative emission develops with a growth rate $g_{\rm SF} \approx \Omega_c^2/(4\alpha) \propto \Delta N$. Here, $\alpha$ is the total field decay rate in a mean-field approximation, including both absorption in the medium and escape of radiation out of the sample. However, this case is not relevant for our samples, which are characterized by a low field dissipation rate within the pumped region, i.e., $\Omega_c /2 > \gamma > \alpha$.  In the latter case, the growth rate saturates at its maximum value $g_{\rm SF} \approx \Omega_c/2 \propto (\Delta N)^{1/2} $.  Note that, in both cases, the SF growth rate is faster than the phase relaxation rate $\gamma$. This ensures a maximally coherent nature of the process.  In a sense, SF establishes an absolute upper limit on the rate with which an ensemble of initially incoherent inverted oscillators can radiate their stored energy.

The dynamics of SF is distinctly different from ASE or lasing.  First, SF develops within a broad spectral bandwidth $\sim g_{\rm SF}> \gamma$ and thus cannot be described by usual rate equations based on adiabatic elimination of the polarization.  Second, after the degenerate (inverted) population of carriers is established on a given LL, the SF pulse is emitted with the delay time $t_{\rm d} \sim (1/g_{\rm SF}) \ln(I_{\rm SF}/I_{0})$, which is logarithmically larger than the inverse growth rate, where the logarithm factor of order 10 is due to the exponential growth of the field from the quantum noise level, $I_{0}$, to the peak intensity, $I_{\rm SF}$. Third, the pulse duration, $\tau_{\rm SF} \sim 2/g_{\rm SF}$, and the delay time decrease with $\Delta N$, and therefore, the pulse intensity $I_{\rm SF} \sim N_{\rm exc}/\tau_{\rm SF}$ scales superlinearly ($\propto N_{\rm exc}^{3/2}$) with the carrier density or the pump pulse energy until all electron states in a given LL are occupied.
 
The coherence length over which an individual pulse is formed is given by $L_c \sim ct_d/\mu$, which is a logarithmic factor of order 10 longer than the exponential amplification length $\sim c/(\mu g_{\rm SF})$.  For a much longer length, the SF will result in multiple pulse formation from independent segments of length $L_c$. For a much shorter length, the SF will not fully develop.   In our system, the magnitude of the cooperative frequency $\Omega_c$ is of the order of 1\,ps$^{-1}$ assuming complete inversion. This translates into a coherence length of the order of 1\,mm.  Therefore, we chose the pumped spot size of the same order, around 0.5\,mm.

To model the dynamics of SF, we first determine appropriate basis states. The important feature of our system is a strong excitonic effect,  which affects both energy levels and matrix elements of the optical transitions. Therefore, we use the exciton basis states taking into account both the magnetic field and the Coulomb interaction on an equal footing; namely, we solved the Schr{\"o}dinger equation for the QW system including an electron and a heavy hole.  The Hamiltonian includes their QW confinement in the growth direction, their motion in the magnetic field, and the Coulomb interaction between them.  A general eigenstate of the Hamiltonian includes mixing among different levels in the growth direction. However, for the first few eigenstates of the Hamiltonian, this kind of mixing is very small, due to the large energy spacing between them. Thus, we can assume that both the electron and hole are in their ground states described by wavefunctions $\phi_{e1}(z_e)$ and $\phi_{h1}(z_h)$ in the growth direction, and their in-plane motion experiences an effective Coulomb interaction given by~\cite{Ivchenko05Book}
\begin{eqnarray}
V_{\rm eff}(\rho) = - \frac{e^2}{\epsilon} \int d z_e \int d z_h \frac{|\phi_{e1}(z_e)|^2 |\phi_{h1}(z_h)|^2}{\sqrt{\rho^2 + (z_e-z_h)^2}}
\label{Coulomb2Deff}
\end{eqnarray}
where $\rho = |\vec{\rho}_e - \vec{\rho}_h|$ is the in-plane electron-hole separation.

Using the symmetric gauge $\vec{A}_{e(h)} = \vec{B} \times \vec{\rho}_{e(h)}/2$, we transform the in-plane electron-hole wavefunction $\Psi(\vec{\rho}_e,\vec{\rho}_h)$ as $U(\vec{\rho})$ = $\exp[-i (e / 2 c \hbar) \vec{B} \cdot (\vec{\rho}_{\rm cm} \times \vec{\rho})] \Psi(\vec{\rho}_e,\vec{\rho}_h)$~\cite{ShinadaTanaka70JPSJ}, where we have set the center-of-mass momentum to be zero, since only these states are relevant to optical transitions. If one writes $U(\vec{\rho})$ = $R_m(\rho) e^{i m \phi} / \sqrt{2\pi}$, where $m$ is the angular momentum number, $R_m(\rho)$ satisfies the equation
\begin{eqnarray}
\left\{ -\frac{\hbar^2}{2\mu_e} \left[ \frac{1}{\rho} \frac{d}{d\rho} \left(\rho \frac{d}{d\rho}\right) - \frac{m^2}{\rho^2} \right] + V_{\rm eff}(\rho) + \frac{\mu_e \omega_c^2 \rho^2}{8} \right.  \nonumber \\
\left. - \frac{m_e-m_h}{m_e+m_h} \frac{m\hbar\omega_c}{2} \right\} R_m(\rho) = \left(E_m - E_{\rm g,eff}\right) R_m(\rho)
\label{EigenEq}
\end{eqnarray}
where $\mu_e$ is the reduced mass, $\omega_c = e B/\mu_e c$, and $E_{\rm g,eff}$ is the effective bandgap, which includes the bandgap of InGaAs and the ground-level confinement energies of the electron and hole. The boundary conditions are $R_m(0)$ being finite and $R_m(\infty) = 0$. This equation has the Sturm-Liouville form but is singular at $\rho = 0$ if $m \neq 0$. It can be solved numerically by using the SLEIGN2 code~\cite{BaileyetAl01ACM}. We calculated the eigen-energies for optical transitions ($m$ = 0), and the results agree with the experimental data shown in Fig.~\ref{setup} extremely well, with deviation less than 1\,meV.  We then proceed with calculating the dipole matrix elements $d_{ii}$ of the interband optical transitions between LLs.  The combination of the magnetic field and Coulomb confinement increases the magnitude of the matrix elements  due to an increased overlap of electron and hole wavefunctions. For example, when $B$ increases from 0 to 17\,T, $|d_{22}|^2$ increases by a factor of 9. Therefore, one should expect that the pulse intensity will increase while the pulse duration and delay time will decrease with increasing magnetic field, in agreement with data presented in Fig.~\ref{fits}.
 
To describe the nonlinear stage of SF pulse formation and propagation, we solved a coupled set of space- and time-dependent density-matrix and wave equations for exciton populations, interband coherences (off-diagonal density-matrix elements), and optical fields.  In the experiment, the carriers were created by a femtosecond pulse above the GaAs barriers and then relaxed down to the lowest subband, occupying the few lowest LLs to the near-complete degeneracy, before the SF pulses were formed.  Therefore, we included in the modeling only the exciton states and recombination radiation pulses corresponding to the (00), (11), and (22) transitions.  The initial kinetics of carrier cooling and relaxation were modeled by an effective scattering rate to the (22) state, which we assumed to be a pulse of amplitude 0.1-1\,ps$^{-1}$ and exponential decay time of order 10\,ps.  The pumping pulse parameters slightly affected the SF delay time but had little effect on the subsequent SF pulse amplitude and duration.  The dominant interaction between excitons that leads to their ultrafast decay and formation of SF pulses is their coupling to a common electromagnetic mode.  Other interactions between excitons were ignored since they are not essential to the SF dynamics, although they renormalize transition energies and Rabi frequencies.
  
\begin{figure}
\begin{center}
\includegraphics[scale=0.5]{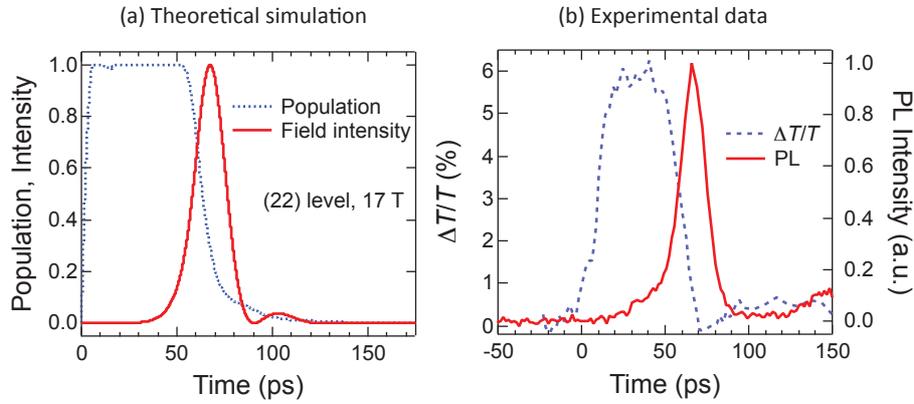}
\caption{(a)~Theoretical simulations of superfluorescence from an ultradense electron-hole plasma in a semiconductor quantum well in a perpendicular magnetic field of 17\,T. Occupation number of excitons on the (22) level (blue dashed line) and normalized electromagnetic field intensity of superfluorescence from the (22) level (red solid line) as a function of time since the beginning of the pump pulse.  Population equal to 1 corresponds to total inversion where all states are fully occupied by excitons.  (b)~Experimental pump-probe (blue dashed line) and TRPL (red solid line) data taken for the (22) level at 17.5\,T and 5\,K.}
\label{comparison}
\end{center}
\end{figure}

The calculated evolution of the exciton population and SF electromagnetic field intensity for the (22) LL at $T$ = 5\,K and $B$ = 17\,T is shown in Fig.~\ref{comparison}(a), together with the corresponding experimental pump-probe and TRPL data in Fig.~\ref{comparison}(b). The simulation shows how the SF pulse depletes all exciton density at the (22) LL, bringing the effective population inversion, $\Delta N$, to $-$1. This is a signature of coherent light-matter interaction, similar to a $\pi$-pulse propagation.  It is important to point out that a pulse of ASE would consume only half of the exciton population, bringing $\Delta N$ to 0, not $-$1. Figure \ref{comparison} demonstrates excellent overall agreement between our simulation and experimental data.  Most importantly, an abrupt disappearance of carriers (sharp edge of the plateau at around 60\,ps time delay) coincides with the radiation pulse detected by the streak camera.  Simulations for the (11) LL give similar results.  
 
The combination of complete initial degeneracy and low temperature makes the relaxation time $T_2$ very long (tens of ps).  In our simulations, we assumed $T_2$ = 50\,ps. An increase in temperature would lead to an increase in the relaxation rate and the transition from SF to a much slower spontaneous emission. This behavior can be seen in Fig.~\ref{PP}, where the sharp edge of the plateau in pump-probe differential transmission morphs into a long tail as the temperature increases from 5 to 150\,K. 
 %
 %
Depending on the pumping rate, simulations also indicate the presence of a second pulse, and sometimes multiple pulses, from the (00) LL. The second pulse results from fast relaxation from higher LLs, which replenishes carrier population depleted by the first pulse. This is consistent with the observed statistics of strong directionality fluctuations between radiation intensities received by two edges, as shown in Fig.~\ref{YD}. Indeed, the data shows that the two fiber outputs are either correlated or anti-correlated in roughly equal proportion.   On average, in 50\% of the shots both edges receive a SF pulse, and in the other 50\% of the shots, only one edge will receive both pulses, in qualitative agreement with observations.
 
\section{Summary and Outlook}

In summary, we have reviewed our recent experimental and theoretical studies to confirm the existence of superfluorescence (SF) in a semiconductor, i.e, cooperative recombination of a large number of electron-hole pairs, resulting in a short burst of coherent, intense, and directional radiation.  This observation not only provides a fully controllable environment in which to study many-body physics but also is promising for future development of novel solid-state sources for producing bursts of coherent radiation.  Namely, the solid-state realization of SF has resulted in an unprecedented degree of controllability in the generation of coherent pulses, opening up opportunities for both fundamental many-body studies and device applications.  We demonstrated that the intensity and delay time of SF bursts are fully tunable through an external magnetic field, temperature, and pump laser power.  Unlike atoms, we can tune virtually everything: the density of states, the oscillator strength, the transition frequencies, and, most importantly, the decoherence times via magnetic field  and temperature.  The fact that a magnetic field affects both the characteristics of SF and the strength of Coulomb interaction can lead to interesting aspects of SF in a way that other systems cannot.  Furthermore, we can make use of advanced semiconductor technology and design compact devices with better modal confinement that can produce SF pulses with desired properties even at weak magnetic fields and room temperature.  Eventually, an electrically-driven device for producing coherent SF pulses with any desired wavelengths can be expected to be developed by utilizing existing technologies of semiconductor quantum engineering.

\section*{Acknowledgement}

This work was supported by the National Science Foundation through grants DMR-1006663 and ECS-0547019. A portion of this work was performed at the National High Magnetic Field Laboratory, supported by NSF Co-operative Agreement No.~DMR-0084173 and by the State of Florida. We thank Dr.~Glenn S.~Solomon for providing us with the InGaAs/GaAs quantum well sample used in this study.

%

\providecommand{\WileyBibTextsc}{}
\let\textsc\WileyBibTextsc
\providecommand{\othercit}{}
\providecommand{\jr}[1]{#1}
\providecommand{\etal}{~et~al.}

\end{document}